\documentclass[11pt]{article}
\begin{document}

\begin{center}
\large \textbf{Weak cosmic censorship conjecture in BTZ black holes with scalar fields}
\end{center}

\begin{center}
Deyou Chen
\end{center}

\begin{center}
School of Science, Xihua University, Chengdu 610039, China

E-mail: deyouchen@hotmail.com
\end{center}

{\bf Abstract:} The weak cosmic censorship conjecture in the near-extremal BTZ black hole has been tested by the test particles and fields. It was claimed that this black hole could be overspun. In this paper, we review the thermodynamics and weak cosmic censorship conjecture in BTZ black holes by the scattering of the scalar field. The first law of thermodynamics in the non-extremal BTZ black hole is recovered. For the extremal and near-extremal black holes, due to the divergence of the variation of the entropy, we test the weak cosmic censorship conjecture by evaluating the minimum values of the function $f$. Both of the extremal and near-extremal black holes cannot be overspun.


\section{Introduction}

It is widely believed that spacetime singularities are formed at the end of gravitational collapses. At the singularities, all physical laws break down. To avoid the destruction caused by the singularities, Penrose proposed the weak cosmic censorship conjecture (WCCC) in 1969 \cite{RP}. This conjecture states that naked singularities cannot be produced in a physical process from regular initial conditions. For black holes, their singularities should be hidden behind event horizons without any access to distant observers. Since the conjecture was put forward, a lot of research has been done on its validity. However, no concrete evidence has been found to prove it and no unanimous conclusion has been reached.

The Gedanken experiment proposed by Wald is the first attempt to test the validity of the WCCC \cite{RW}. In this experiment, a test particle with energy, large enough charge and angular momentum is thrown into a Kerr-Newman black hole to test whether the black hole exceeds its extremal limit. The Kerr-Newman solution
describes a charged and rotating spacetime. Its charge and angular momentum per unit mass are bounded by the mass as $a^2+Q^2\leq M^2$. When $M^2< a^2+Q^2$, the black hole exceeds its extremal limit and the event horizon disappears. Thus the singularity is naked and the WCCC is violated. Following this seminal work, people investigated the validity from the aspects of particles and fields. Jacobson and Sotiriou studied the absorption of an object with spin and orbital angular momentum in a near-extremal Kerr black hole. They found that the black hole could be overspun without consideration of the radiative and self-force effects \cite{JS}. The overcharge of the near-extremal Reissner-Nordstrom black hole was first found in \cite{VEH} and then investigated with consideration of the tunnelling effects in \cite{MS}. However, when the radiative, backreaction and self-force effects are taken into account, particles may be escaped from black holes and naked singularises can be avoided \cite{SH,BCK1,BCK2,ZVPH,IST,CB,CBSM}. This result was confirmed again in the recent work where the self-force and finite size effects were considered \cite{RMW2,SW}. A counterexample to the WCCC in four-dimensional anti de Sitter (AdS) spacetime was presented in \cite{CS}. In this spacetime, constant time slices have planar topology. It was shown that it is just a pure AdS in the past. In the future, the curvature grows without bound and leaves regions of spacetime with arbitrarily large curvatures. These regions are naked to boundary observers. This work is important to the weak gravity conjecture \cite{HMNV}.

The validity of the WCCC was researched from the aspect of fields \cite{DS,GZT,IS,BG3,KD}. In the research, the field has a finite energy, which indicates the existence of the wave packet. Initially,  the field does not exist and there is only a black hole. The field comes in from infinity and interacts with the black hole. Due to the interaction, the energy, charge and angular momentum are transferred between the field and the black hole. Part of the field is reflected back to infinity. Finally, the field decays away leaving behind another spacetime with the new energy, charge and angular momentum \cite{GZT}. Whether the black hole is overspun or overcharged can be judged by the change of event horizon. Based on this view, the interaction between a dyonic Kerr-Newman black hole and a complex massive scalar field was discussed \cite{IS}. It was found that this interaction did not destroy the WCCC. The same result was derived by Toth \cite{GZT}. His derivation is based on a null energy condition, conservation laws and the electromagnetic field of the black hole. In the recent work, Gwak calculated the variations of the energy and angular momentum of the Kerr-(Anti-)de Sitter black hole during a infinitesimal time interval by the fluxes of energy and angular momentum of the scalar field \cite{BG3}. He found that the black hole kept the initial states and was not overspun. This result is different from that of the near-extremal black hole by throwing a test particle into it. When initial data are non-generic, naked singularities can be developed \cite{JMC1,JMC2,JMC3,JMC4,JMC5,JMC6,JMC7}. Other tests of the WCCC are referred to \cite{HS,RV,SH2,KD2,GZ,CHS,MST,BG1,BG2,GBG,JCRO,DJ,LWL,LCNR,RS,GMZZ,NQV,ASZZ,YW,CYZ1,CYZ2,KD,WWY1,MT1,HMT1,HOP1,ZHH2} and the references in them.

The Ba$\tilde{n}$ados, Teitelboim and Zanelli (BTZ) black hole is a solution of Einstein field equations in three-dimensional spacetime and describes a rotating AdS geometry \cite{BTZ,BHTZ}. The related properties of BTZ black holes were researched in \cite{KD,RC,WSYY,MOP}. In \cite{RC}, Rocha and Cardoso tested the WCCC in the extremal BTZ black hole by using the test particle. They found that this black hole could not be overspun. In this work, the initial black hole was designated as an extremal one. However, D$\ddot{u}$ztas claimed that a near-extremal black hole had the possibility to exceed its extremal limit whether or not the superradiation occurred \cite{KD}. If the superradiation did not occur, overspinning becomes generic and also applies to the extremal BTZ black hole. If there was the superradiation, the black hole is overspun in a certain range of frequencies. He elaborated this result from the massive test particles and the test fields, respectively. Due to the similarities of rotating black holes, the same result was gotten in the Kerr black hole \cite{DS}. Obviously, this result is different from that derived by Wald and Sorce \cite{RMW2,SW}.

In this paper, we review the thermodynamics and WCCC in BTZ black holes by the scattering of a scalar field. The BTZ black holes are generic in the sense of constituting an open set in the space of solutions to the Einstein equation. The change of the energy and angular momentum of the black hole during a time interval relies on the fluxes of energy and angular momentum of the ingoing wave function. To get the wave function, we introduce a tortoise coordinate and solve the second-order differential equation at the event horizon. Here the time interval is infinitesimal. For the non-extremal BTZ black hole, the increase of the event horizon ensures that the WCCC is valid. The first law of thermodynamics is recovered by the scattering. For the near-extremal and extremal BTZ black holes, Eq.(\ref{eq3.15}) is divergent at their event horizons. Therefore, we need to resort to other methods to test the validity. The validity is tested by evaluating the minimum value of the function $f$. It is found that their horizons do not disappear and the singularities are always hidden behind the horizons.

The rest is organized as follows. The BTZ black hole solution is given and its thermodynamics are discussed in the next section. In section 3, the first law of thermodynamics in the non-extremal BTZ black hole is recovered by the scattering of the scalar field. In section 4, the validity of the WCCC in the extremal and near-extremal BTZ black holes is tested by the minimum values of the function $f$. Section 5 is devoted to our discussion and conclusion.

\section{BTZ black holes}

The BTZ metric is given by \cite{BTZ,BHTZ}

\begin{eqnarray}
ds^{2} & =&  -f dt^2+\frac{1}{f} dr^{2} +r^2(d\varphi-\frac{J}{2r^2}dt)^2,
\label{eq2.1}
\end{eqnarray}

\noindent where

\begin{eqnarray}
f &=& f(M,J,r)= -M+ \frac{r^2}{l^2} + \frac{J^2}{4r^2}.
\label{eq2.2}
\end{eqnarray}

\noindent It describes a locally three-dimensional rotating AdS spacetime. The parameter $l^2$ is related to the cosmological constant $\Lambda$ as $l^2 = -\frac{1}{\Lambda}$.  $M$ and $J$ are the ADM mass and angular momentum, respectively. They determine the asymptotic behavior of the spacetime. The event (inner) horizons locate at $r_+(r_-)$ and satisfy the relations

\begin{eqnarray}
Ml^2=r_+^2+r_-^2, \quad\quad J^2l^2=4r_+^2r_-^2.
\end{eqnarray}

\noindent When $r_+=r_-$, the two horizons are coincident with each other and the black hole becomes an extremal one.

The entropy, Hawking temperature, angular velocity and ADM mass are

\begin{eqnarray}
S &=& 4\pi r_+, \quad\quad T =\frac{r_+}{2\pi l^2}-\frac{J^2}{8\pi r_+^3}, \nonumber\\
\Omega &=& \frac{J}{2r_+^2}, \quad \quad M= \frac{r_+^2}{l^2}+\frac{J^2}{4r_+^2},
\label{eq2.3}
\end{eqnarray}

\noindent respectively. Here, the expression of the entropy used in \cite{BTZ} is adopted, which shows that the entropy is equal to twice the perimeter length of the horizon. When the mass of a BTZ black hole varies, other thermodynamic quantities of the black hole, such as entropy, temperature and angular velocity, and so on, will also vary. These thermodynamic quantities obey the first law of thermodynamics

\begin{eqnarray}
dM = TdS+\Omega dJ.
\label{eq2.4}
\end{eqnarray}

\noindent We will find that the first law is recovered by the scattering of the scalar field in the next section. These variations are caused by the interaction between the scalar field and the black hole. Due to the interaction, the energy and angular momentum are transferred and they are evaluated by the energy flux and angular momentum flux. Therefore, we first write the action and calculate the energy-momentum tensor.

\section{Thermodynamics of non-extremal BTZ black holes}

The action for the minimally coupled complex scalar field under the BTZ spacetime is

\begin{eqnarray}
I= \int{dtdrd\varphi\sqrt{-g}\mathcal{L}}= -\frac{1}{2}\int{dtdrd\varphi\sqrt{-g}[\partial_{\mu} \Phi \partial^{\mu} \Phi^{\star} + \mu_0^2\Phi\Phi^{\star}]},
\label{eq3.1}
\end{eqnarray}

\noindent where $\mathcal{L}$ is the Lagrangian density and $\mu_0$ is the mass \cite{KKS}. Then the energy-momentum tensor is obtained from the action, namely, $T_{\mu\nu}=-\frac{2}{\sqrt{-g}}\frac{\delta I}{\delta g^{\mu\nu}}$. We get

\begin{eqnarray}
T_{\nu}^{\mu} = \frac{1}{2}\partial^{\mu} \Phi \partial_{\nu} \Phi^{\star} + \frac{1}{2}\partial^{\mu} \Phi^{\star} \partial_{\nu} \Phi +\delta_{\nu}^{\mu}\mathcal{L}.
\label{eq3.2}
\end{eqnarray}

\noindent To evaluate the energy-momentum tensor, we need to know the expression of the wave equation $\Phi$ which obeys the equation of motion for the scalar field. This equation is obtained from the action (\ref{eq3.1}) and takes on the form

\begin{eqnarray}
\frac{1}{\sqrt{-g}}\frac{\partial}{\partial x^{\mu}}\left(\sqrt{-g}g^{\mu\nu}\frac{\partial \Phi}{\partial x^{\nu}}\right)-\mu_0^2\Phi = 0.
\label{eq3.3}
\end{eqnarray}

Due to the existence of the killing vectors $(\frac{\partial}{\partial t})^a$ and $(\frac{\partial}{\partial \varphi})^a$ in the BTZ background spacetime, we carry out the separation of variables

\begin{eqnarray}
\Phi = e^{-i(\omega t - j \varphi)}R(r),
\label{eq3.4}
\end{eqnarray}

\noindent where $\omega$ and $j$ denote the energy and angular momentum, respectively. Inserting the contravariant components of the BTZ metric and the separation of variables (\ref{eq3.4}) into the Klein-Gordon equation (\ref{eq3.3}) yields the second-order differential equation. To solve this equation, we introduce a tortoise coordinate \cite{HL}

\begin{eqnarray}
r_{\star}=r+\frac{1}{2\kappa}ln\frac{r-r_+}{r_+},
\label{eq3.5}
\end{eqnarray}

\noindent where $\kappa=2\pi T$ is the surface gravity at the event horizon and $T$ is the Hawking temperature. Then the second-order differential equation becomes

\begin{eqnarray}
\frac{d^2R(r)}{dr_{\star}^2}&+&\frac{f^2+f(2\kappa r+1)}{(f+1)^2r}\frac{dR(r)}{dr_{\star}}  \nonumber\\
&+&\frac{4\omega^2r^4-4\omega Jjr^2+4j^2r^2f-J^2j^2-4f\mu_0^2r^4}{4(f+1)^2r^4}R(r)=0.
\label{eq3.6}
\end{eqnarray}

\noindent Since we are interested in the scattering near the event horizon, the equation is going to be solved at the horizon. Let $r\rightarrow r_+$. Eq.(\ref{eq3.6}) is reduced to

\begin{eqnarray}
\frac{d^2R(r)}{dr_{\star}^2}+(\omega - j\Omega)R(r)=0.
\label{eq3.7}
\end{eqnarray}

\noindent In the above equation, $\Omega = \frac{J}{2r_+^2}$ is the angular velocity at the event horizon. From Eq.(\ref{eq3.7}), we get the radial solutions

\begin{eqnarray}
R(r)\sim e^{\pm i (\omega -j\Omega)r_{\star}},
\label{eq3.8}
\end{eqnarray}

\noindent where the solutions with $+(-)$ denote the outgoing (ingoing) radial waves. Therefore, the standard wave equations are

\begin{eqnarray}
\Phi = e^{-i(\omega t - j \varphi)\pm i (\omega -j\Omega)r_{\star}}.
\label{eq3.9}
\end{eqnarray}

\noindent The interaction between the field and the black hole transfers the energy and angular momentum. Since we are discussing the horizon changes after the black hole absorbs the energy and angular momentum, we focus our attention on the ingoing wave equation.

Two Killing vectors  $(\frac{\partial}{\partial t})^a$ and $(\frac{\partial}{\partial \varphi})^a$ correspond to two local conservation laws in the BTZ spacetime. The corresponding conservative quantities are energy $E$ and angular momentum $L$. When the fluxes of energy and angular momentum flow into the event horizon and are absorbed by the black hole, the energy and angular momentum of the hole change. The energy flux and angular momentum flux are

\begin{eqnarray}
\frac{dE}{dt}=\int{T_{t}^r \sqrt{-g}d\varphi}, \quad \frac{d L}{dt}=-\int{T_{\varphi}^r \sqrt{-g}d\varphi},
\label{eq3.10}
\end{eqnarray}

\noindent respectively. Combining the fluxes with the energy-momentum tensor and the ingoing wave equation yields

\begin{eqnarray}
\frac{dE}{dt}=2\pi r_+\omega(\omega-j\Omega), \quad \frac{d L}{dt}=2\pi r_+j(\omega-j\Omega).
\label{eq3.11}
\end{eqnarray}

\noindent In the derivation, $\frac{d r_{\star}}{d r}=\frac{f+1}{f}$ obtained from Eq.(\ref{eq3.5}) was used. The energy of the black hole is its ADM mass and the angular momentum is expressed as $J$. Therefore, the increases of the energy and angular momentum during the time interval $dt$ are

\begin{eqnarray}
dM= 2\pi r_+\omega(\omega-j\Omega)dt, \quad dJ= 2\pi r_+j(\omega-j\Omega)dt,
\label{eq3.12}
\end{eqnarray}

\noindent which may be negative or positive values depending on the sign of the value of $\omega - j\Omega$. When $\omega > j\Omega$, the energy and angular momentum of the black hole increase and the fluxes flow into the event horizon. There is no change of the energy and angular momentum for $\omega = j\Omega$. $\omega < j\Omega$ implies the decrease of the energy and angular momentum. Then the energy and angular momentum are extracted by the scattering and the superradiation occurs. In fact, the appearance of superradiation should satisfy that the boundary condition of the scalar field is in the asymptotic region. Here we follow the work of Gwak and focus on the infinitesimal change in the BTZ spacetime \cite{BG3}. Therefore, our discussion does not rely on the asymptotic boundary conditions.

In the following discussion, the time interval is assumed to be infinitesimal. Correspondingly, the variations of the energy and angular momentum are also infinitesimal. The scattering varies the function $f$ and the horizon radius $r_+$. The variations are labeled as $\delta f$ and $d r_+$, respectively. These variations satisfy

\begin{eqnarray}
\delta f&=&f(M+dM,J+dJ,r_++dr_+)-f(M,J,r_+) \nonumber\\
&=& \left.\frac{\partial f(M,J,r)}{\partial M}\right|_{r=r_+}dM+ \left.\frac{\partial f(M,J,r)}{\partial J}\right|_{r=r_+}dJ+ \left.\frac{\partial f(M,J,r)}{\partial r}\right|_{r=r_+}dr_+,
\label{eq3.13}
\end{eqnarray}

\noindent where

\begin{eqnarray}
\left.\frac{\partial f(M,J,r)}{\partial M}\right|_{r=r_+}&=&-1, \quad \quad  \left.\frac{\partial f(M,J,r)}{\partial J}\right|_{r=r_+}= \frac{J}{2r_+^2}, \nonumber\\  \left.\frac{\partial f(M,J,r)}{\partial r}\right|_{r=r_+}&=& 4\pi T.
\label{eq3.13-1}
\end{eqnarray}

\noindent To derive $dr_+$, one can assume that the final state is still a black hole after the absorption of the infinitesimal energy and angular momentum \cite{BG1,BG3}. This implies $f(M+dM,J+dJ,r_++dr_+)=f(M,J,r_+)= 0$. Thus, the variation of the horizon radius is

\begin{eqnarray}
dr_+=\frac{ r_+(\omega - j\Omega)^2dt}{2T}.
\label{eq3.14}
\end{eqnarray}

\noindent When $\omega \neq j\Omega$, we get $dr_+>0$. When $\omega = j\Omega$, we have $dr_+=0$. Therefore, the horizon radius does not decrease when the black hole absorbs the ingoing wave. It implies that the singularity is hidden behind the event horizon and can not be observed by external observers of the black hole. Using the relation between the entropy and horizon radius, we get

\begin{eqnarray}
dS=\frac{ 2\pi r_+(\omega - j\Omega)^2dt}{T}.
\label{eq3.15}
\end{eqnarray}

\noindent It shows that the entropy does not decrease under the scattering of the field. This result supports the second law of thermodynamics and is a simple
consequence of the fact that the system satisfies the null energy condition. From Eqs.(\ref{eq3.12}) and (\ref{eq3.15}), we get

\begin{eqnarray}
dM = TdS+\Omega dJ.
\label{eq3.16}
\end{eqnarray}

\noindent Therefore, the first law of thermodynamics in the non-extremal BTZ black hole is recovered by the scattering of the scalar field.

In the thought experiment, people usually prefer to study systems which inferentially close to the critical condition. In the next section, we will investigate this case, namely, near-extremal and extremal BTZ black holes. For these black holes, Eq.(\ref{eq3.15}) is divergent at their event horizons. Thus, the above method can not be applied to the extremal and near-extremal BTZ black holes. We need to resort to other methods to test the WCCC.

\section{WCCC in near-extremal and extremal BTZ black holes}

The WCCC in the near-extremal and extremal BTZ black holes has been tested. It was found that the near-extremal BTZ black hole had the possibility to be overspun \cite{KD}. But the extremal BTZ black hole could not be overspun \cite{RC}. In this section, we review the validity of the WCCC in the near-extremal and extremal BTZ black holes by the minimum values of the function $f$ at the final states. Due to the interaction between the black hole and the field, the energy and angular momentum of the black hole change. Correspondingly, the value of the function $f$ changes. In the metric, there are two roots (correspond to the inner and event horizons, respectively) for $f<0$ and a root (corresponds to the event horizon) for $f=0$. When $f>0$, the event horizon disappears and the singularity is naked.

In this section, the time interval is also infinitesimal. Correspondingly, the transferred energy and angular momentum via the scattering is  also infinitesimal. The minimum value of $f$ is expressed as $f_0=f(M,J,r_0)=-M +\frac{r_0^2}{l^2}+\frac{J^2}{4r_0^2}$, where $r_0$ is the location corresponding to $f_0$. $r_0$ is not an independent variable, depending on $M$ and $J$. Thus we get

\begin{eqnarray}
&& f(M+dM,J+dJ,r_0+dr_0)\nonumber\\
&=& f_0+ \left.\frac{\partial f(M,J,r)}{\partial M}\right|_{r=r_0}dM+ \left.\frac{\partial f(M,J,r)}{\partial J}\right|_{r=r_0}dJ+ \left.\frac{\partial f(M,J,r)}{\partial r}\right|_{r=r_0}dr_0 \nonumber\\
&=& -\left(\frac{\omega}{j}\right)^2 2\pi j^2r_+dt+\left(\frac{\omega}{j}\right)2\pi j^2r_+\left(\Omega+\Omega_0\right)dt +f_0-2\pi j^2r_+\Omega\Omega_0 dt,
\label{eq4.1}
\end{eqnarray}

\noindent where

\begin{eqnarray}
\left.\frac{\partial f(M,J,r)}{\partial M}\right|_{r=r_0}=-1, \quad \left.\frac{\partial f(M,J,r)}{\partial J}\right|_{r=r_0}= \frac{J}{2r_0^2}, \quad \left.\frac{\partial f(M,J,r)}{\partial r}\right|_{r=r_0}=0,
\label{eq4.2}
\end{eqnarray}

\noindent and $\Omega_0=\frac{J}{2r_0^2}$ is the angular velocity at the location $r_0$. The formulae in Eq.(\ref{eq3.12}) was used to derive Eq.(\ref{eq4.1}). The above equation is a quadratic equation about $\frac{\omega}{j}$ and has a maximal value by adjusting the value of $\frac{\omega}{j}$. If this maximal value is greater than zero, there is no horizon. Instead, the event horizons exist.

For the extremal BTZ black hole, the event and inner horizons are coincident with each other and the temperature is zero. Thus, the term $TdS$ in Eq.(\ref{eq2.4}) disappears and $dM=\Omega dJ$. Using Eq.(\ref{eq3.12}), we can easily get $\omega=j\Omega$. The location of the event horizon is coincident with that of the minimum value of the function $f$, namely, $r_0=r_+$. Thus, Eq.(\ref{eq4.1}) is written as

\begin{eqnarray}
f(M+dM,J+dJ,r_0+dr_0) =  -2\pi r_+\left(\omega-j\Omega\right)^2dt=0.
\label{eq4.3}
\end{eqnarray}

\noindent This result shows that the extremal black hole is also extremal one with the new mass and angular momentum under the scattering. Therefore, the extremal BTZ black hole cannot be overspun. This result is full in accordance with that gotten by Rocha and Cardoso in \cite{RC}, where the WCCC is tested by throwing a point particle into the black hole.

For the near extremal BTZ black hole, there are $f_0 < 0$ and $|f_0|\ll 1$. To get the maximal value of the function, we order $r_+=r_0 +\epsilon$, where $0<\epsilon\ll1$. Meanwhile, we let $dt$ be on the infinitesimal scale and $\epsilon \sim dt$. Thus, the function $f_0$ is simplified to

\begin{eqnarray}
f_0 = -\frac{2r_+}{l^2}\epsilon+\frac{J^2}{2r_+^3}\epsilon <0.
\label{eq4.5}
\end{eqnarray}

\noindent For convenience of discussion, we rewrite Eq.(\ref{eq4.1}) as a function about $\frac{\omega}{j}$ and get

\begin{eqnarray}
f\left(\frac{\omega}{j}\right)&=& -2\pi j^2r_+\epsilon\left(\frac{\omega}{j}\right)^2 \nonumber\\
&& +2\pi j^2r_+(\Omega+\Omega_0)\epsilon \left(\frac{\omega}{j}\right) -\frac{2r_+}{l^2}\epsilon+\frac{J^2}{2r_+^3}\epsilon -2\pi j^2r_+\Omega\Omega_0\epsilon.
\label{eq4.6}
\end{eqnarray}

\noindent The maximum value, which locates at $\frac{\omega}{j}=\frac{\Omega+\Omega_0}{2}$, is

\begin{eqnarray}
f\left(\frac{\omega}{j}\right)_{max} = -\frac{2r_+}{l^2}\epsilon+\frac{J^2}{2r_+^3}\epsilon+ \mathcal{O}(\epsilon),
\label{eq4.7}
\end{eqnarray}

\noindent where $\mathcal{O}(\epsilon)=\frac{2\pi J^2j^2}{r_+^5}\epsilon^3$ can be neglected. Using Eq.(\ref{eq4.5}), we find $f\left(\frac{\omega}{j}\right)_{max}<0$. This implies that there are two roots existed in the function. Therefore, the event and inner horizons do not disappear under the scattering of the scalar field and the singularity is hidden behind the event horizon. In \cite{KD}, D$\ddot{u}$ztas found that the near-extremal BTZ black hole could be overspun via the research of the particle absorption and field effects. Clearly, our result is different from that gotten by him.

\section{Discussion and Conclusion}

In this paper, we investigated the thermodynamics and WCCC in the BTZ black holes by the scattering of the scalar field. The variations of the energy and angular momentum in the black holes during an infinitesimal time interval were calculated. The first law of thermodynamics in the non-extremal BTZ black hole was recovered by the scattering. The increase of the horizon radius ensures that the singularity is hidden behind the event horizon of the non-extremal black hole. For the near-extremal and extremal BTZ black holes, since Eq.(\ref{eq3.15}) is divergent, we tested the WCCC by evaluating the minimum values of the function $f$ at the final states. We found that these two black holes maintain their near-extremity and extremity respectively. This result is full consistence with that gotten by Wald and Jorse \cite{RMW2,SW}.

In the recent work, D$\ddot{u}$ztas directly assumed that the horizons were destroyed and obtained the relationship between the frequency and azimuthal wave number of the incident field \cite{KD}. When the superradiation occurred, this frequency range satisfied $\frac{Jn}{Ml^2(1+2\epsilon)}<\omega<\frac{n}{l(1+\epsilon)}$, where $n$ is the azimuthal wave number. When there was no superradiation, it obeyed $0<\omega<\frac{n}{l(1+\epsilon)}$. For the extremal BTZ black hole, he found that it couldn't overspin if the supperradiation occurred. However, if there was no superradiation for the field, overspinning would appear. In our investigation, we did not directly assume that the black hole horizons are destroyed. We found that the extremal and near-extremal BTZ black holes can not be overspun in any frequencies range.

\textbf{Acknowledgments}
This work is supported by the National Natural Science Foundation of China (Grant nos. 11205125, 11875095).


\end{document}